\documentclass[twocolumn,showpacs,amsmath,amssymb]{revtex4}

\usepackage{graphicx}
\usepackage{dcolumn}
\usepackage{bm}
\usepackage{fixme}
\usepackage{mathtools}

\DeclarePairedDelimiter\abs{\lvert}{\rvert}  
\DeclarePairedDelimiter\bra{\langle}{\rvert} 
\DeclarePairedDelimiter\ket{\lvert}{\rangle} 

\renewcommand{\phi}{\varphi}
\renewcommand{\epsilon}{\varepsilon}

\newcommand\diff[3][\partial]{\frac{#1 #2}{#1 #3}}
\newcommand\vek[1]{\mathbf{#1}} 
\newcommand\hop{\mathcal{H}} 

\begin{document}

\title{Honeycomb optical lattices with harmonic confinement}

\author{J. Kusk Block}
\affiliation{Department of Physics and Astronomy, University of Aarhus, DK-8000 {\AA}rhus C, Denmark}
\author{N. Nygaard}
\affiliation{Lundbeck Foundation Theoretical Center for Quantum System Research, Department of Physics and Astronomy, University of Aarhus, DK-8000 {\AA}rhus C, Denmark}

\date{\today}

\begin{abstract}
We consider the fate of the Dirac points in the spectrum of a honeycomb optical lattice in the presence of a harmonic confining potential. By numerically solving the tight binding model we calculate the density of states, and find that the energy dependence can be understood from analytical arguments. In addition, we show that the density of states of the harmonically trapped lattice system can be understood by application of a local density approximation based on the density of states of the homogeneous lattice. The Dirac points are found to survive locally in the trap as evidenced by the local density of states. They furthermore give rise to a distinct spatial profile of a noninteracting Fermi gas.
\end{abstract}

\maketitle

\section{\label{sec:introduction} Introduction}

Graphene is a carbon monolayer with a honeycomb crystal structure, which was only recently produced~\cite{Novoselov2004}. 
The band structure of graphene is intriguing in that the dispersion is linear in the vicinity of the Fermi energy. This makes the material a zero-gap semiconductor, with quasiparticles behaving as massless Dirac fermions, thus opening the possibility of studying quantum electrodynamics with electrons in a solid state system~\cite{Katsnelson20073}. The existence of carriers described by the Dirac equation has been confirmed experimentally along with the demonstration of an anomalous quantum Hall effect~\cite{Novoselov2005,Zhang2005}.
The striking electronic properties of graphene makes it an interesting system not only for studying fundamental physics, but also as a platform for device fabrication~\cite{Geim2007,Neto2009}. 

Building on the potential of graphene as a test bed for relativistic quantum theory, several theoretical papers have pointed out that ultracold atoms in a honeycomb optical lattice could prove an attractive, alternative system for simulating relativistic physics~\cite{Zhu2007,wu:235107,wu:070401,1367-2630-10-10-103027,Haddad20091413,PhysRevA.80.043411,bermudez-2009}. 
An optical lattice is a periodic potential, formed by interfering laser beams, in which atoms exhibit the same Bloch band physics as solid state electrons. But contrary to a solid state crystal both the depth and the geometry of an optical lattice potential can be controlled by adjusting the intensity and configuration of the lasers. Hence an optical lattice provides a pristine environment for implementing condensed matter models, and probing many-body dynamics such as the superfluid to Mott insulator transition~\cite{Greiner2002,Bloch2008}. In addition, from a quantum simulator point of view ultracold atoms posses the advantageous qualities of controllable interactions (using a magnetic field tunable Feshbach resonance~\cite{RevModPhys.78.1311}) and the possibility of mapping the rich internal state space of the atoms onto multiple spin degrees of freedom. By applying additional light fields an artificial gauge field can be engineered~\cite{Lin2009,stanescu:053639}. With non-Abelian gauge fields different topological phases can be engineered~\cite{bermudez-2009}. Other schemes for producing a relativistic dispersion with optical fields have also been proposed~\cite{PhysRevLett.100.130402,PhysRevA.79.043621,PhysRevA.80.063603,PhysRevLett.103.035301}

However, in experiments the discrete translational symmetry of the optical lattice is broken by a trapping potential, which confines the atoms. This complicates the comparison with solid state phenomena, but it is often surmised that the confining potential is slowly varying, and that a local density approximation can therefore be used.

\begin{figure}[htbp]
\begin{center}
\includegraphics[width=\columnwidth]{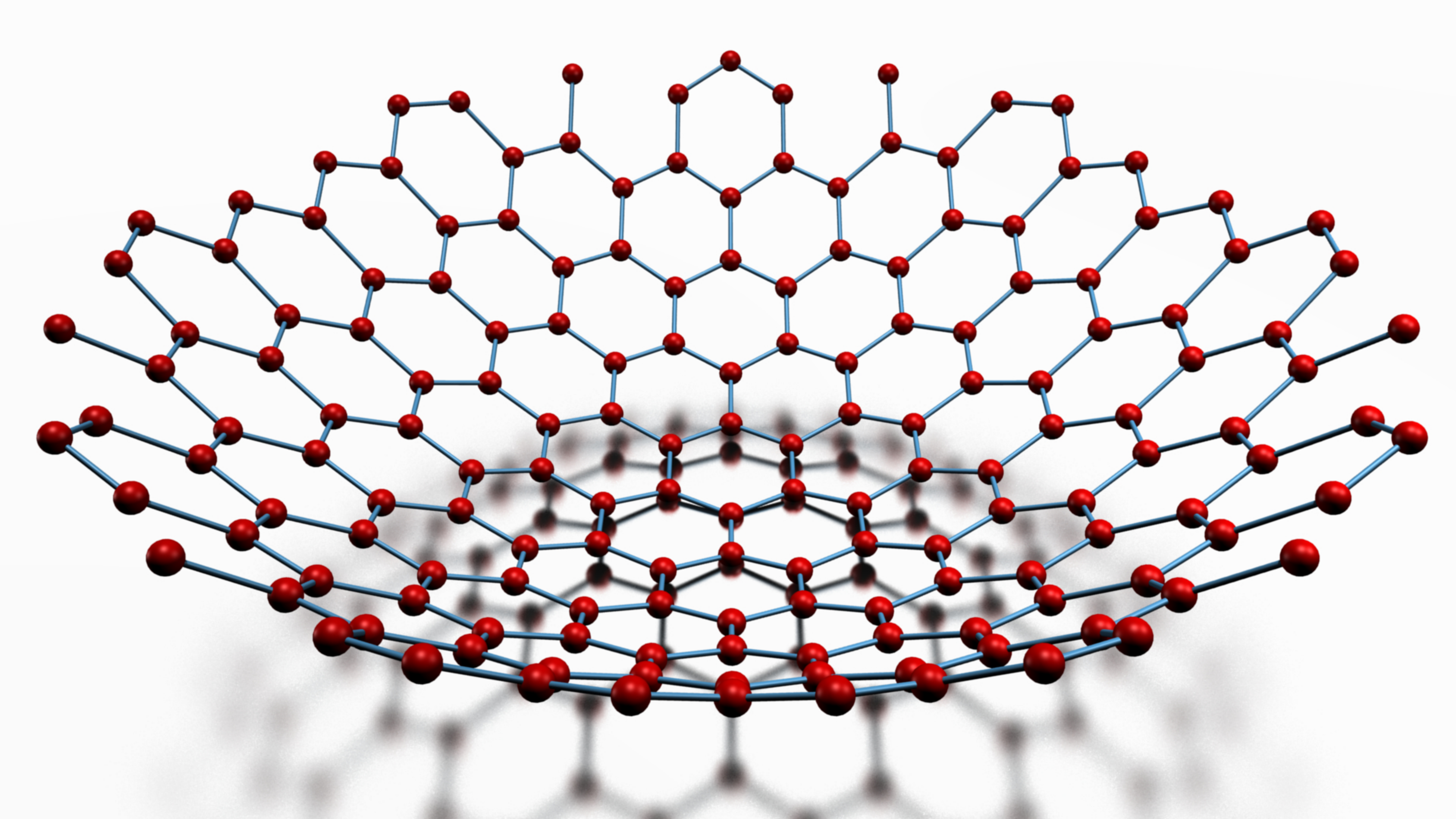} 
\caption{(Color online) Schematic view of the honeycomb lattice with a superposed harmonic trapping potential. The harmonic potential is centered on one of the lattice sites.}
\label{Fig:Schematic}
\end{center}
\end{figure}

In this paper we test the validity of this presumption and determine how the physics of graphene is modified in an inhomogeneous honeycomb optical lattice. Similar calculations have been done for cubic lattices in one and two dimensions~\cite{Hooley2004,Rigol2004,PhysRevLett.93.110401,PhysRevLett.93.120407,PhysRevA.72.033616}. In the present work we first give a brief review of how a honeycomb lattice potential can be generated in an experiment (for a longer discussion see e.g. \cite{PhysRevA.80.043411}). Then we consider the situation illustrated in Fig.~\ref{Fig:Schematic} where the translational symmetry of the lattice is broken by a parabolic offset of the site energies. We solve a tight binding model numerically for a finite system and characterize the spectrum by the density of states. The spectral features can be understood by a combination of analytical arguments and a local density approximation. We address the existence of Dirac particles in the inhomogeneous lattice by plotting the local density of states and by calculating the density distribution of a noninteracting Fermi gas in the combined lattice and harmonic potential.

\section{\label{sec:lattice} Constructing a honeycomb optical lattice}

A honeycomb optical lattice can be constructed by superposing three laser beams with wave vectors $\vek{k}_i$ ($i=1,2,3$) of identical magnitude $k_L=2\pi/\lambda_L$ lying in the $x$-$y$ plane at $2\pi/3$ angles with each other. If the three lasers have the same intensity and are linearly polarized in the $z$-direction, this gives rise to a lattice potential of the form \cite{Blakie2004,PhysRevA.80.043411}
\begin{equation}
V_L(\vek{r})=V_0 \left[ \cos(\vek{b_1}\cdot \vek{r})+\cos(\vek{b_2}\cdot \vek{r})+\cos((\vek{b_1}+\vek{b_2})\cdot \vek{r})\right], 
\label{eq:vlat}
\end{equation}
where $\vek{b_1}=\vek{k}_3-\vek{k}_1,\vek{b_2}=\vek{k}_1-\vek{k}_2$
are the reciprocal lattice vectors. The lattice depth $V_0$ depends on the intensity and the detuning of the lattice lasers, and here we only consider $V_0>0$ corresponding to a positive detuning, which produces the honeycomb structure of graphene with a spacing between nearest neighbor lattice sites of $a_0=2\lambda_L/\sqrt{27}$. A negative laser detuning generates a triangular lattice potential.  
Using additional confinement along the $z$-axis an effectively two-dimensional system can be realized. 


The honeycomb optical lattice described above can be generalized in two straightforward ways. First, if the direction of polarization of the lasers is changed from perpendicular to the lattice plane to coplanar with the wave vectors of the beams, the resulting periodic light field is circularly polarized at the positions of the lattice minima, with the lattice sites forming an alternating hexagonal pattern of $\sigma^+$ and $\sigma^-$ polarizations~\cite{PhysRevLett.70.2249}. In such a lattice atoms in different internal spin states will experience different light shifts~\cite{PhysRevLett.91.010407}, and for atoms with spin projection $|m_F|>0$ the lattice potential becomes a periodic array of offset double wells~\cite{Becker2009}. Secondly, if the laser intensities differ, an anisotropic honeycomb lattice is generated where the tunneling rates depend on direction. If the intensity imbalance is sufficiently large this induces a band gap in the single-particle spectrum, equivalent to the Dirac fermions acquiring mass~\cite{Zhu2007,1367-2630-10-10-103027}.  

In the following we restrict our attention to the spin-independent, isotropic honeycomb lattice. However, the form of the lattice potential above assumes that the three lasers beams are plane waves, while in reality their cross sections have a gaussian intensity profile. This gives rise to an energy offset between different lattice wells, which for a sample much smaller than the beam widths can be approximated by a harmonic oscillator potential. In experiments an additional confining potential is often added intentionally to restrict the size of the cloud. Our motivation for this work is to investigate how the presence of such a spatially dependent energy offset between lattice sites affects the single-particle physics of the honeycomb lattice. 
  


\section{\label{sec:tight_bind} Tight binding model}

We consider a tight binding model with nearest neighbor tunneling and
expand the Hamiltonian in terms of the localized (orthogonal) Wannier
states of the first Bloch band,
\begin{align}
  \hop=-J\sum_{\langle jj'\rangle} \ket{w_j} \bra{w_{j'}}  + \frac{1}{2} \kappa \sum_{j}
  r_j^2 \ket{w_{j}}\bra{w_{j}}, \label{eq:hamho} 
\end{align}
where $\ket{w_j}$ is the Wannier state localized at lattice site
$j$, and $r_j$ is the distance of site $j$ to the center of the trap, which has spring constant $\kappa$. The sum in the first term is over nearest neighbor sites.
The nearest neighbor tunneling amplitude between sites $j$ and $j'$ is defined as
\begin{equation}
  \label{eq:Jdef}
  J=-\bra{w_{j'}}\hat{T}+\hat{V}_L\ket{w_{j}}
\end{equation}
where $\hat{T}$ is the kinetic energy operator in $xy$-plane. Tunneling to next-nearest neighbor sites is strongly suppressed. The tight binding model is illustrated in Fig.~\ref{Fig:Schematic}.
For simplicity we take the center of the trap to coincide with one of the lattice sites. This restriction is easy to relax.

We assume that the harmonic potential does not modify the nearest neighbor tunneling rate. This approximation is valid provided two neighboring wells (at  distances $r$ and $r+\delta r$ from the trap center) are separated by a barrier $V_0$, which is much larger than the energy difference between their minima $\delta E(r)$. For $r\gg a_0$ the energy difference between neighboring points is $\delta E(r)\leq \frac{1}{2}\kappa[(r+a_0)^2-r^2]\approx \kappa a_0r$. This defines an energy cutoff in our model, since $\delta E(r)$ increases with $r$. Thus high energy states with a wave function, which remains finite beyond a critical distance  $r_c=V_0/\kappa a_0$, will not be represented accurately in our model.  Hence we are limited to consider energies $E\ll E_c=\frac{1}{2}\kappa r_c^2-3J$, where $-3J$ is the lowest energy in the spectrum for a homogeneous lattice (see below). The relevant energy scale is set by the tunneling, and we thus require $\kappa a_0^2/J \ll  (V_0/J)^2$. If we introduce the characteristic length scale of the harmonic oscillator $a_{\rm{osc}}=\sqrt{J/\kappa}$ this criterion translates into $a_0/a_{\rm{osc}}\ll V_0/J$.

The oscillator length scale is typically of the order of micrometers, while the lattice lasers have wave lengths of several hundred nm. Since $V_0/J\approx (V_0/E_R)^{1/4}\exp[1.582 \sqrt{V_0/E_R}]$, where $E_R=h^2/2m\lambda_L^2$ is the recoil energy of the lattice lasers~\cite{PhysRevA.80.043411}, the condition for the validity of the tight binding model is almost always satisfied for lattices deeper than about $5 E_R$. 

In the numerical diagonalization we impose hard wall boundary
conditions at $r_{\rm{max}}=60a_0$. This artificial restriction leads
to finite size effects, such as edge states, that may be interesting
in their own right~\cite{stanescu:053639}. Below we also give analytic results, which apply
for an infinite lattice. 



\subsection{Homogeneous lattice dispersion}
\label{sec:homogeneous} 

\begin{figure}[htbp]
\begin{center}
\includegraphics[width=\columnwidth]{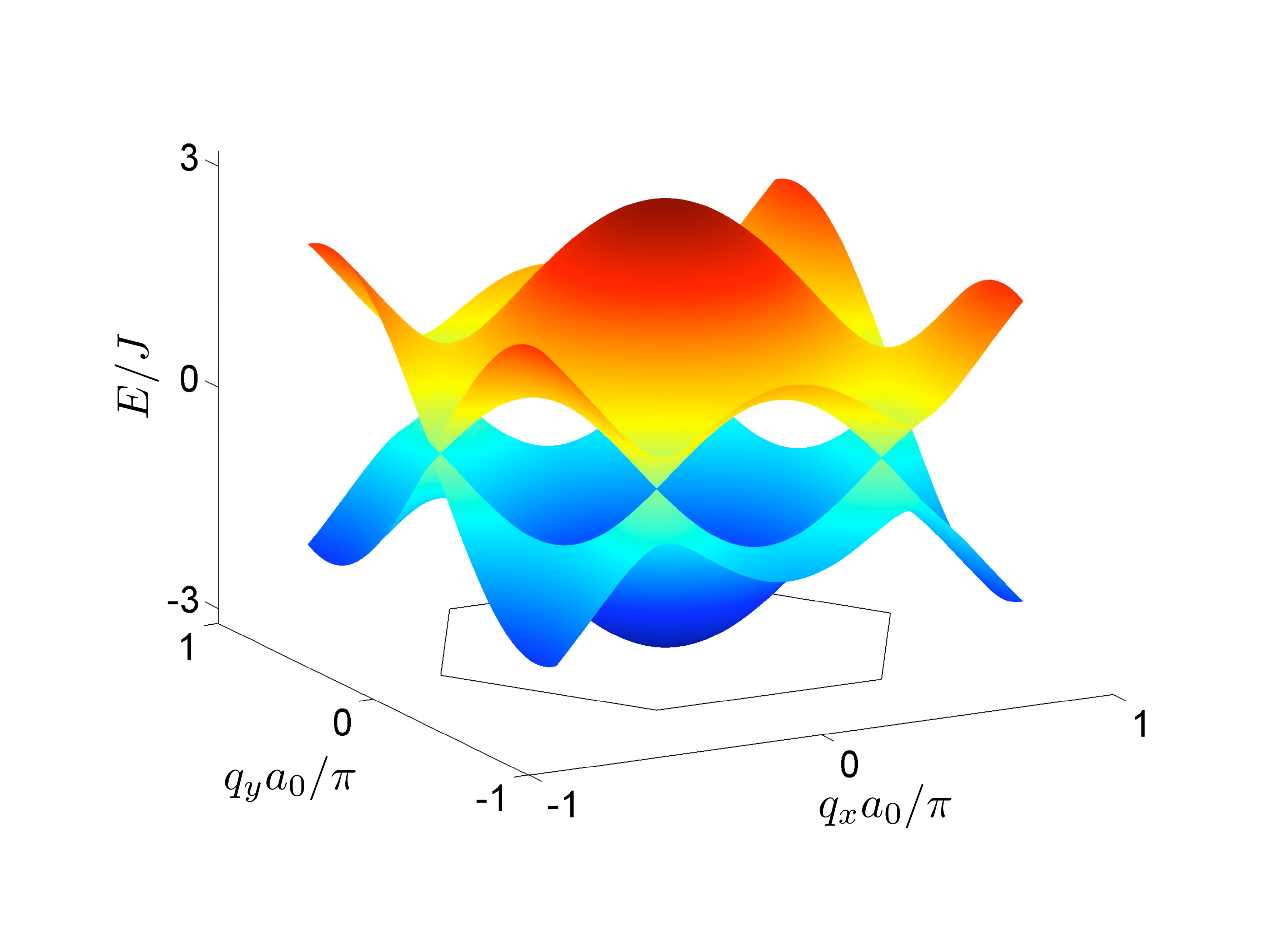}
\caption{(Color online) The single-particle spectrum for an infinite honeycomb lattice. The two bands touch at the six Dirac points located at the edges of the first Brillouin zone, which is indicated by the hexagon in the plane $E=-3J$.}
\label{Fig:Eq_hom}
\end{center}
\end{figure}

We first give a brief review of the homogeneous lattice case where $\kappa=0$. For an infinite lattice the eigenstates of the tight binding Hamiltonian are Bloch waves with energies 
\begin{eqnarray}
E^\pm_{\bf q}&=&\pm J\biggl[3+2\cos(\sqrt{3} q_ya_0) \nonumber \\
&& +4\cos\left(\frac{3}{2} q_xa_0\right)\cos\left(\frac{\sqrt{3}}{2} q_ya_0\right)\biggr]^{1/2}
\label{Eq:Eq}
\end{eqnarray}
as a function of the quasimomentum ${\bf q}$. The spectrum consists of a lower and an uppper band as depicted in Fig.~\ref{Fig:Eq_hom} with a hexagonal first Brillouin zone.  Near the six corners of the first Brillouin zone the two bands form opposing cones, which exactly touch at the corner points. Since each of the corners is shared equally between three adjoining Brillouin zones the first Brillouin zone contains two independent corner points at quasimomenta ${\bf K}$ and ${\bf K}'$. Around these points the dispersion is linear: $E^\pm_{\bf k}\approx \pm \hbar v_{\rm{F}}|{\bf k}|$ for ${\bf q}={\bf K}+{\bf k}$ with $|{\bf k}|\ll |{\bf K}|$ and similarly in the vicinity of ${\bf K}'$. Since this corresponds to the dispersion of massless Dirac fermions with $v_{\rm{F}}=3Ja_0/2\hbar$ playing the role of the speed of light $c$, the quasimomenta ${\bf K}$ and ${\bf K}'$ are referred to as Dirac points~\cite{PhysRevLett.53.2449,Neto2009}. A lot of the excitement about graphene can be attributed to the promise of observing relativistic effects with solid state electrons. While for graphene $v_{\rm{F}}\simeq c/300$ the effective speed of light for atomic Dirac fermions in an optical lattice would typically be of the order of mm/s. 

\section{Single particle Density of states}
\label{DOS}

We now turn to the fate of the Dirac points when a harmonic confining potential is added to the lattice. With the discrete translational symmetry broken, we can expect to find both delocalized states with a well defined quasimomentum and localized states consisting of many quasimomentum components. It is therefore no longer meaningful to discuss the dispersion, and instead we look for evidence of the Dirac points in the single-particle density of states (DOS)
\begin{equation}
\rho(E) = \sum_n \delta(E-E_n).
\label{Eq:DOS_def}
\end{equation}
Here the sum is over the eigenstates of the tight binding Hamiltonian $\hop|\psi_n\rangle=E_n|\psi_n\rangle$. Numerically, we find $\rho(E)$ by binning the eigenvalues into small energy intervals of varying width. Counting the number of eigenstates in each interval gives a good approximation to the DOS in the middle of the intervals, provided the widths of the intervals are small enough to capture the variation of $\rho(E)$ with energy, but large enough that fluctuations are smeared out. 

Before investigating the DOS in the inhomogeneous lattice we first recall how the Dirac points are manifested in the form of the DOS in the absence of the trap. Since an analytic expression for $\rho(E)$ exists for the infinite lattice, this also constitutes a test of the numerics.  

\subsection{Homogeneous lattice}
\label{sec:DOS_hom}

For the homogeneous lattice the single-particle DOS per unit cell 
has the analytical form~\cite{hobson:662,Neto2009} 
\begin{equation}
\rho_0(E) = \frac{2}{\pi^2}\frac{|E|}{J^2}\frac{1}{\sqrt{Z_0}}K\left(\sqrt{\frac{Z_1}{Z_0}}\right),
\end{equation}
where $K(z)=\int_0^{\pi/2}[1-z\sin^2t]^{-1/2}dt$ is the complete elliptic integral of the first kind, and 
\begin{eqnarray}
Z_0 &=& \left\{ 
\begin{array}{ll}
(1+|\frac{E}{J}|)^2-\frac{[(E/J)^2-1]^2}{4} & , \ |\frac{E}{J}|\leq 1\\
4|\frac{E}{J}| & , \  1 \leq |\frac{E}{J}| \leq 3
\end{array}   
\right. 
\\
Z_1 &=& \left\{ 
\begin{array}{ll}
4|\frac{E}{J}|& , \ |\frac{E}{J}|\leq 1\\
(1+|\frac{E}{J}|)^2-\frac{[(E/J)^2-1]^2}{4}  & , \  1 \leq |\frac{E}{J}| \leq 3
\end{array}   
\right. 
\label{Eq:DOS_hom}
\end{eqnarray}

\begin{figure}[htbp]
\begin{center}
\includegraphics[width=\columnwidth]{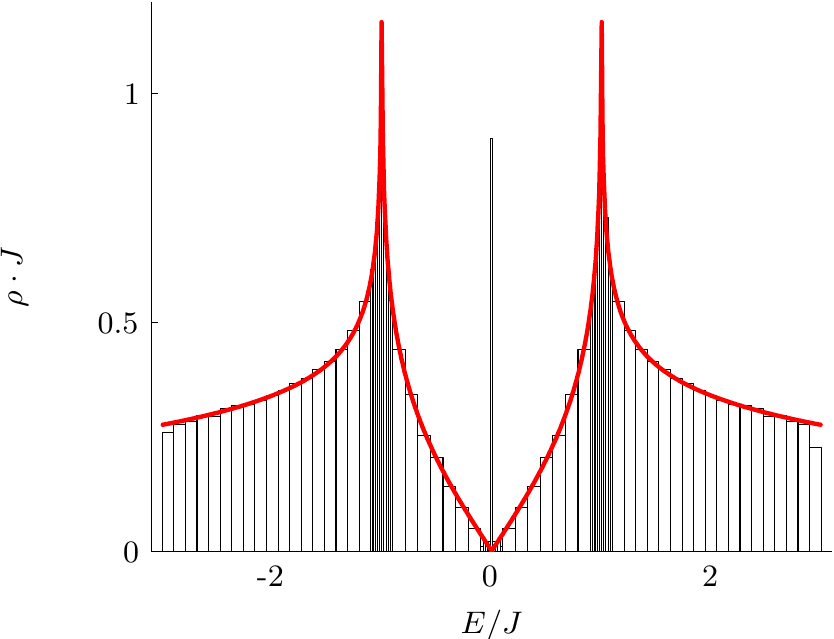}  
\caption{(Color online) Single-particle density of states per unit cell for the
  homogeneous lattice. The solid line is the analytical expression for
  an infinite lattice, (\ref{Eq:DOS_hom}), while the histogram is the
  binned density of states from the numerical calculation with
  $r_{max}=60a_0$, corresponding to 8792 lattice sites. The bin size
  is varied to resolve the details in the spectrum.} 
\label{Fig:DOS_hom}
\end{center}
\end{figure}

The analytical DOS is plotted in Fig.~\ref{Fig:DOS_hom}. In the vicinity of the Dirac point ($E=0$) the linear dispersion leads to a DOS which vanishes as $\rho_0(E)\propto|E|$ with no band gap. The van Hove singularities at $E=\pm J$ arise due to the saddle points in the single-particle spectrum at the edge of the Brillouin zone, halfway between neighboring Dirac points. We note that the DOS is symmetric around the Dirac point, $\rho_0(-E)=\rho_0(E)$. 
The spectral symmetry is broken if next-nearest-neighbor tunneling is included in the tight binding Hamiltonian. 

The histogram in Fig.~\ref{Fig:DOS_hom} is the numerically calculated DOS, which agrees with the analytical expression except for a large peak at $E=0$. This additional peak is due to edge states, an artifact of our finite numerical grid. These zero energy modes are localized at the boundary of the system and appear because we confine the system in a cylindrical box. But they can be studied in graphene nanoribbons~\cite{Neto2009} and could be constructed in an optical lattice by applying a repulsive potential at the edge of the cloud~\cite{stanescu:053639}.    

\subsection{Inhomogeneous lattice}
\label{sec:DOS_inhom}

\begin{figure*}[htb]
\begin{minipage}{0.49\textwidth} 
\includegraphics[width=\textwidth,height=0.22\textheight,keepaspectratio]{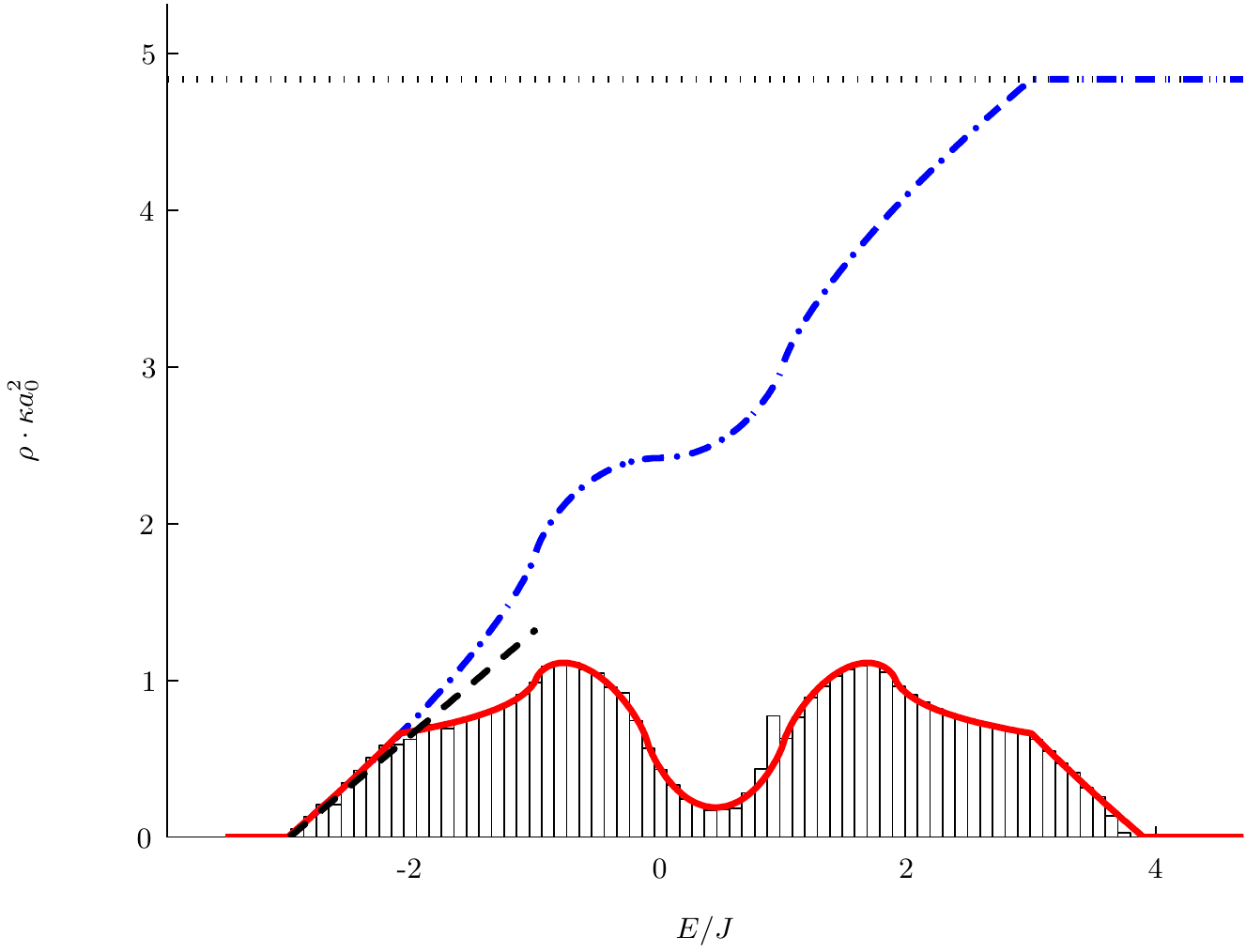} 
\end{minipage}
\begin{minipage}{0.49\textwidth}
\includegraphics[width=\textwidth,height=0.22\textheight,keepaspectratio]{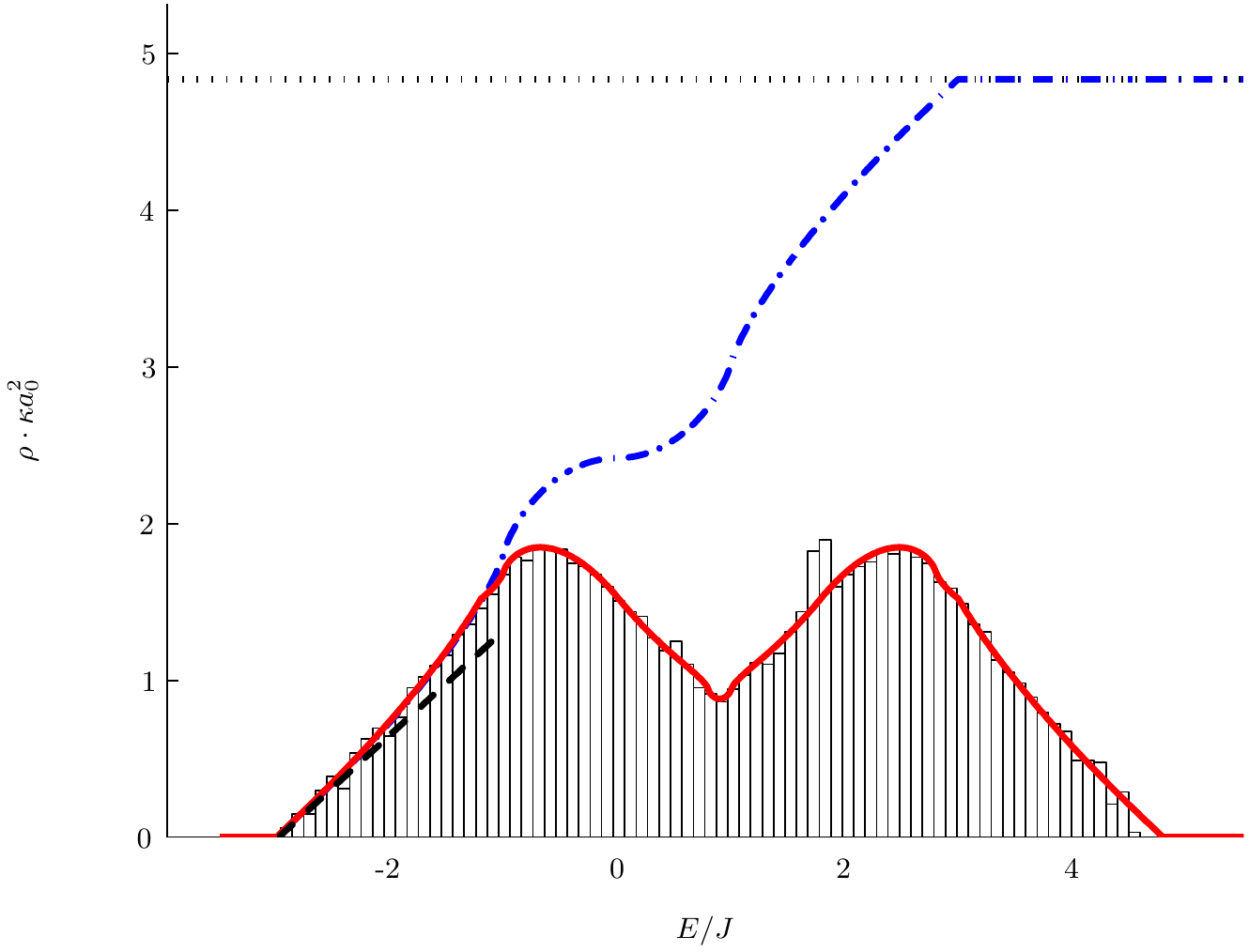}
\end{minipage}
\begin{minipage}{0.49\textwidth}
\includegraphics[width=\textwidth,height=0.22\textheight,keepaspectratio]{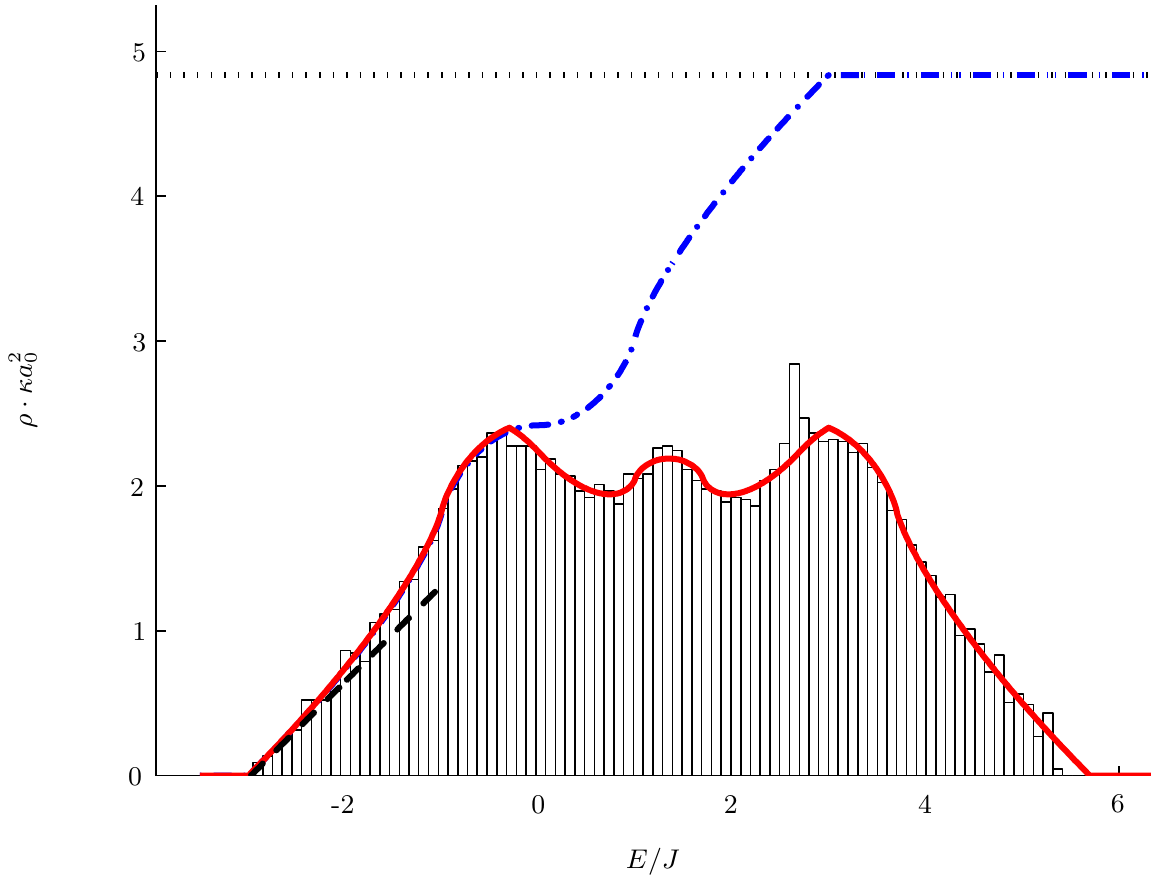}
\end{minipage}
\begin{minipage}{0.49\textwidth}
\includegraphics[width=\textwidth,height=0.22\textheight,keepaspectratio]{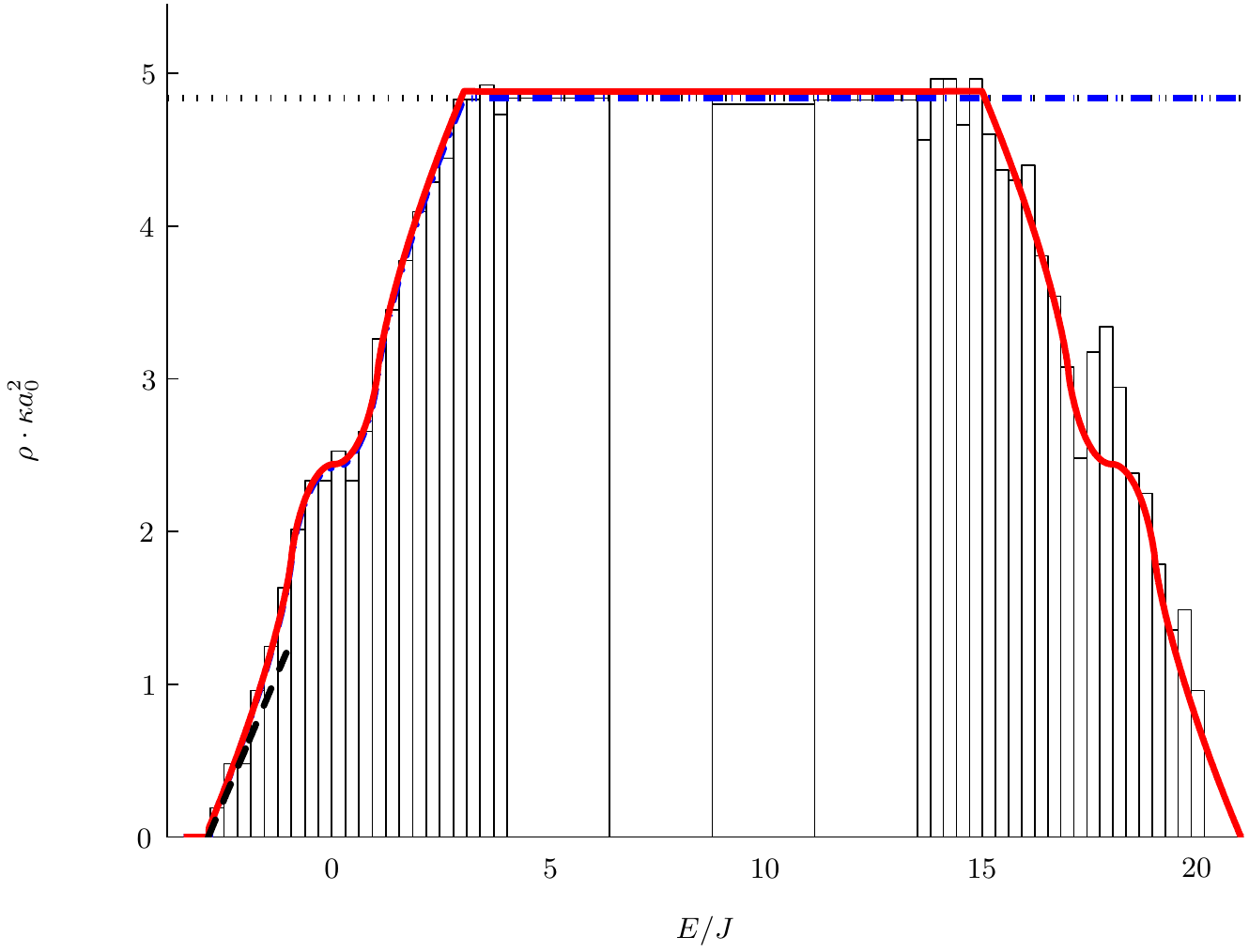}
\end{minipage}
\caption{(Color online) Binned density of states for four different
  trap strengths $\kappa  a_0^2/J=5\cdot 10^{-4},1\cdot 10^{-3},1.5\cdot 10^{-3}$ and
  $1\cdot 10^{-2}$ (histograms). The solid line is calculated using a
  local density approximation for $\rho(E)$ (see text). The dashed and the dotted
  lines are the low energy limit (\ref{eq:lowdos}) and the high energy plateau (\ref{eq:doshigh}), respectively, while the
  dash-dotted line represents the local density approximation for
  $\rho(E)$ on a lattice confined in an infinite box.}
\label{Fig:DOS_inhom}
\end{figure*}

We now turn to the combined lattice and harmonic trapping potential. In Fig.~\ref{Fig:DOS_inhom} we plot the binned density of states for a range of trap strengths. 
We make the following observations on the form of the DOS of the finite system: as the trap strength is increased from zero the characteristic valley around the Dirac point at $E=0$ is gradually filled in, and the minimum is shifted to higher energies. For $\kappa r_{\rm{max}}^2>12J$ ($\kappa a_0^2>3.3\cdot 10^{-3}J$ for $r_{\rm{max}}=60a_0$) the valley has been replaced by a plateau, and as $\kappa$ is increased further the length of this plateau is extended. 
The peak due to the edge states is shifted to $E=\frac{1}{2}\kappa r_{\rm{max}}^2$ , as expected for eigenstates localized at the edge of the cylindrical box. At the same time the peak is broadened due to mixing of the localized edge states with delocalized states in the same energy range. 
Lastly, the symmetry of the DOS is observed to be nearly conserved (apart from the edge state feature), but around an energy $E_0=\frac{1}{4}\kappa r_{\rm{max}}^2> 0$, such that $\rho(E_0-E)=\rho(E_0+E)$. Below we explain each of these observations by analytic arguments.


\subsubsection{Low energy limit}
\label{sec:DOS_inhom_lowE}

In the low energy limit, the lower band of the pure lattice
has a dispersion resembling that of a free particle with an effective mass, $m^*=\hbar^2(\partial^2 E_{\bf q}/\partial q_x^2|_{q=0})^{-1}=2\hbar^2/3Ja_0^2$. Hence the low energy DOS is that of a 2D harmonic oscillator, $\rho(E)=(E-E_{\rm{min}})/(\hbar \omega^*)^2$, with
a characteristic frequency $\omega^*=\sqrt{\kappa/m^*}$ and a minimum energy $E_{\rm{min}}=-3J+\hbar \omega^*$ given by the infimum of the lattice spectrum offset by the zero point energy of the oscillator. The low energy DOS is therefore a linear function of the energy for $E>E_{\rm{min}}$:
\begin{equation}
\rho(E)=\frac{2}{3 \kappa a_0^2}\bigg(\frac{E}{J}+3-\sqrt{\frac{\kappa
    a_0^2}{2J}} \bigg).
  \label{eq:lowdos}
\end{equation} 


\subsubsection{High energy limit}
\label{sec:DOS_inhom_lowE}

At high energies $E \gg J$ the kinetic (and lattice) energy 
is negligible compared with the trap energy and \eqref{eq:hamho} reduces
to the potential energy of a 2D harmonic oscillator. 
The eigenstates of the trap potential energy operator are localized states with
energies $E_j = \frac{1}{2}\kappa \abs{\vek{r}_{j}}^2$. These have been observed experimentally in a one-dimensional optical lattice with harmonic confinement~\cite{PhysRevLett.93.120407}. The
DOS is then
\begin{equation}
  \label{eq:doshighdef}
  \rho(E)=\diff[d]{N(E)}{E}=\diff[d]{N(r)}{r} \diff[d]{r}{E},
\end{equation}
where $N(E)$ is the number of quantum states with energy less than $E$ and $N(r)$ is the number of lattice points in a circle of radius
$r$. Geometric considerations show that the DOS in the high
energy limit approaches the constant value
\begin{equation}
  \label{eq:doshigh}
  \rho(E)=\frac{8 \pi  }{3\sqrt{3}} \frac{1}{\kappa a_0^2}.
\end{equation}
Accordingly, $\rho \kappa a_0^2$ forms a plateau at $8\pi/3\sqrt{3}\simeq 4.84$ at high energies as affirmed by the numerical spectrum in~Fig.~\ref{Fig:DOS_inhom}  

In the finite system the plateau in the DOS is observed to begin at $E=3J$ and end at $E=\frac{1}{2}\kappa r_{\rm{max}}^2-3J$. Hence the plateau appears if $\kappa r_{\rm{max}}^2>12 J$. At higher energies the DOS decreases with increasing energy ultimately vanishing at the largest eigenvalue in the spectrum, which is approximately given by $E_{\rm{max}}=\frac{1}{2}\kappa r_{\rm{max}}^2+3J$. These observations are explained in section~\ref{sec:DOS_inhom_LDA} below where we discuss an approximation to the spectrum based on the slow variation of the trapping potential on the scale of the lattice modulation.



\section{Local density of states}
\label{sec:Local_DOS}

While the Dirac point in the global DOS is erased by adding a confining potential to the lattice we now investigate if it survives locally by calculating the local density of states (LDOS), which is indicative of the local structure of the spectrum. Specifically, we calculate the angle-averaged LDOS
\begin{equation}
\rho(E,r) = \sum_n \int_0^{2\pi} \frac{d\phi}{2\pi} \, |\psi_n({\bf r})|^2\delta(E-E_n),
\label{eq:LDOS}
\end{equation}  
by a binning procedure, where we add the probability densities of all eigenstates in a narrow interval of both $E$ and $r$.

\begin{figure}[htbp]
\begin{center}
   \includegraphics[width=\columnwidth]{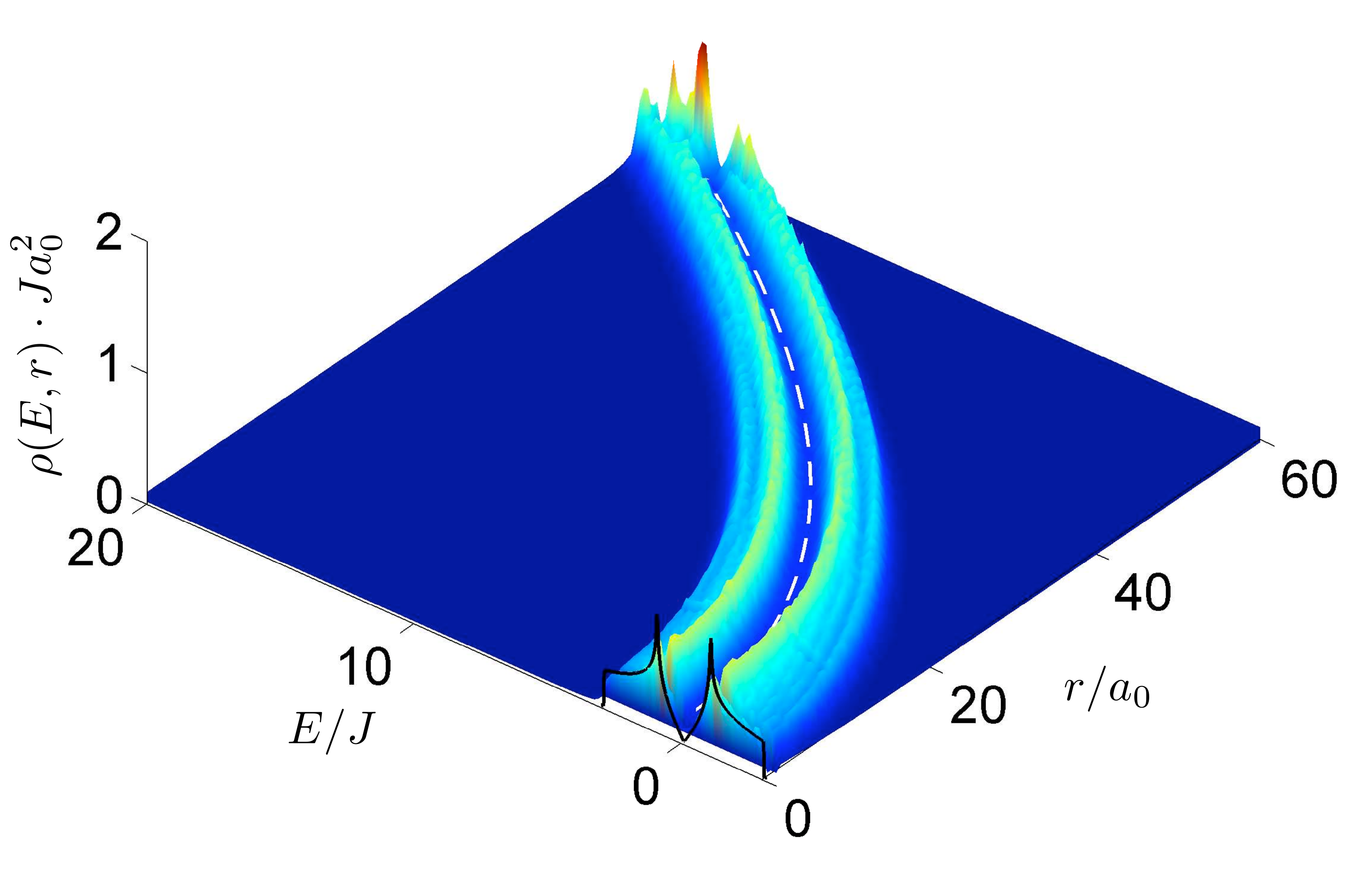}
\caption{(Color online) The local density of states as a function of energy and distance from the center of the trap for $\kappa a_0^2=0.01J$. For clarity the increase in $\rho(E,r)$ due to the expanding number of lattice sites in the enclosed area as $r$ increases has been removed. The large peak at $r=r_{\rm{max}}$ is due to the edge states. The harmonic oscillator potential energy is indicated by the dashed line. The analytical DOS for a homogeneous lattice is indicated at $r=0$.}
\label{fig:LDOS}
\end{center}
\end{figure}

As is clear from Fig.~\ref{fig:LDOS} the LDOS as a function of energy at a fixed $r$ looks just like a local copy of the homogeneous lattice DOS displaced along the energy axis by the local harmonic potential energy $\frac{1}{2}\kappa r^2$ with the edge states visible as a large peak at $r=r_{\rm{max}}$ and $E=\frac{1}{2}\kappa r_{\rm{max}}^2$. By our averaging procedure the van Hove singularities are rounded. This demonstrates that the Dirac physics of graphene is accessible in an inhomogeneous honeycomb lattice, provided local spectroscopic probes are available. For clarity the local density of states has been divided by $4\pi r \Delta r/(3\sqrt{3})$, which is the number of lattice sites in a radial shell between $r$ and $r+\Delta r$.

\section{Local density approximation}
\label{sec:DOS_inhom_LDA}

If $a_{\rm{osc}}\gg a_0$ there is no appreciable change in the harmonic potential over several units cells, and an approximation where the lattice is taken to be locally homogeneous can be expected to be good. With this and the suggestive form of $\rho(E,r)$ in mind we now construct a local density approximation (LDA) to gain further insight into the shape of the DOS. Semi-classically the local DOS for a unit cell at the distance $r$ from the trap center is given by
\begin{equation}
\rho_{LDA}(E,r) = \sum_{\bf q} \delta(E-E_{\bf q}(r)),
\end{equation}
where $E_{\bf q}(r)=E_{\bf q}+\frac{1}{2}\kappa r^2$. This is just the DOS for the homogeneous lattice shifted by the local harmonic potential energy, {\textit{i.e.}} $\rho_{LDA}(E,r)=\rho_0(E-\frac{1}{2}\kappa r^2)$. In the LDA the global DOS is found by integrating $\rho_{LDA}(E,r)$ over the entire lattice, weighted by the number of lattice sites at each distance $r$. This is proportional to $2\pi r$, and the DOS may therefore be approximated by
\begin{equation}
\rho_{LDA}(E) = \frac{2\pi{\mathcal{N}}}{\kappa} \int_{u_1}^{u_2}  \rho_0\left(u \right)  du.
\label{eq:LDA_DOS}
\end{equation}
The normalization constant ${\mathcal{N}}$ is chosen such that $\int_{-\infty}^{\infty} \rho_{LDA}(E)dE$ gives the total number of lattice sites inside the radius $r_{\rm{max}}$, and we have substituted $u=E-\frac{1}{2}\kappa r^2$.
The finite support of the homogeneous lattice DOS implies the lower and upper limits $u_1=\max(-3J,E-\frac{1}{2}\kappa r_{\rm{max}}^2)$ and $u_2=\min(3J,E)$, respectively.

This LDA is plotted in Fig.~\ref{Fig:DOS_inhom} and shows a remarkable agreement with the numerically calculated DOS, apart from the edge states, which are not captured by the semi-classical estimate. The efficacy of the LDA was demonstrated for a three-dimensional cubic lattice with a harmonic confining potential in \cite{baillie:033620}, and the method should be valid in any optical lattice potential as long as the condition $a_{\rm{osc}}\gg a_0$ is satisfied. 

Based on the semi-classical estimate we can explain the following features of the DOS: 

{\textit{Scaling}}: in (\ref{eq:LDA_DOS}) the integral only depends on the strength of the trapping potential through the lower limit $u_1$. Hence if $E<\frac{1}{2}\kappa r_{\rm{max}}^2-3J$ the value of the integral is only a function of $E$. This implies a universal form of $\kappa \rho_{LDA}(E)$ in the limit of an infinite lattice such that $\rho_{LDA}\propto \kappa^{-1}$ for all energies. This agrees with (\ref{eq:lowdos}) (in the limit where $\kappa a_0^2\ll  J$ such that the LDA is valid) and with (\ref{eq:doshigh}). The universal form of $\kappa \rho_{LDA}(E)$ for an infinite lattice is indicated by the dashed-dotted line in Fig.~\ref{Fig:DOS_inhom}. 

{\textit{Limits}}: $\rho_{LDA}$ vanishes for $E<-3J$, consistent with the analytical low energy estimate $E_{\rm{min}}$ above, when the zero-point energy of the trap can be neglected. The semi-classical DOS also vanishes at energies $E>E_{\rm{max}}$.

{\textit{Plateau}}: the high energy plateau is also characterized by considering the limits in (\ref{eq:LDA_DOS}). If $3J<E<\frac{1}{2}\kappa r_{\rm{max}}^2-3J$ the integral is over the entire homogeneous lattice DOS and equals a constant independent of $E$. Therefore $\rho_{LDA}(E)=(2\pi/\kappa)\times{\rm{const}}$. in that case. This explains the beginning and the end of the plateau. The condition for the plateau to appear is $3J<\frac{1}{2}\kappa r_{\rm{max}}^2-3J$ or $\kappa r_{\rm{max}}^2>12J$.

{\textit{Symmetry}}: within the LDA we can understand the symmetry of the DOS as follows: if we neglect the small zero point energy $\hbar\omega^*$ the center of the spectrum is given by $E_0=(E_{\rm{max}}+E_{\rm{min}})/2=\frac{1}{4}\kappa r_{\rm{max}}^2$. By another change of variable to $v=E_0\pm E-\frac{1}{2}\kappa r^2$ the DOS at $E_0\pm E$ can then be written as 
\begin{equation}
\rho_{LDA}(E_0 \pm E) = \frac{2\pi{\mathcal{N}}}{\kappa} \int_{-E_0\pm E}^{E_0 \pm E} \,  \rho_0(v)dv.
\end{equation}
By partitioning the integration interval the integral can be split into two part $\rho_{LDA}=\rho^I_{LDA}+\rho^{II}_{LDA}$, where the first part 
\begin{equation}
\rho^I_{LDA}(E_0 \pm E) = \frac{2\pi{\mathcal{N}}}{\kappa} \int_{-E_0+E}^{E_0 -E} \,  \rho_0(v)dv.
\end{equation}
is the same for both arguments. For simplicity we consider only $E>0$. The second part  is
\begin{equation}
\rho^{II}_{LDA}(E_0 \pm E) = \frac{2\pi{\mathcal{N}}}{\kappa} \int_{\pm E_0 - E}^{\pm E_0 +E} \,  \rho_0(v)dv.
\end{equation}
Since the homogeneous lattice DOS is symmetric about zero energy, $\rho_0(-E)=\rho_0(E)$, it follows that $\rho_{LDA}(E_0-E)=\rho_{LDA}(E_0+E)$.

It is important to stress that the symmetry of the DOS for the trapped system is a finite size effect. The same applies for the critical value of $\kappa$ for the onset of the high energy plateau as well as the finite length of the plateau as a function of energy for a fixed trap strength.
For an unbounded system the high energy plateau stretches to infinitely high energies and $\kappa \rho(E)$ follows a universal form as discussed above. This is shown by the dashed-dotted line in Fig.~\ref{Fig:DOS_inhom}, which represent $\rho_{LDA}(E)$ in the limit where $r_{\rm{max}}\gg \sqrt{2(E+3J)/\kappa}$, such that finite size effects are irrelevant for the energies shown (note that $E\ll V_0^2/2\kappa a_0^2-3J$ is needed for the tight binding model to be applicable, c.f. Section~\ref{sec:tight_bind}).  It is worth noting that while the DOS for the finite system develops gradually from that of the homogeneous lattice as the trapping strength is increased from zero, the DOS of the trapped, unbounded system is {\textit{qualitatively}} different from its translationally invariant counterpart, owing to the divergence of the harmonic oscillator potential as $r\rightarrow \infty$. This dramatic difference between the infinite system DOS for $\kappa=0$ and in the limit $\kappa \rightarrow 0$ was also noted by Hooley and Quintanilla for a cubic lattice~\cite{Hooley2004}.

\section{Fermionic density profile}
\label{sec:Density}

Above we have accounted for the spectrum of a single atom in honeycomb lattice with harmonic confinement. We have found that the Dirac points of the homogeneous graphene spectrum survive locally in the presence of the harmonic trapping potential. In this section we consider how this can be confirmed experimentally. While Bragg scattering has been applied with great success as a spectroscopic probe of atomic quantum gases~\cite{PhysRevLett.82.871,PhysRevLett.82.4569,Enst2009}, a calculation of the response of a many-body system to this kind of perturbation is beyond the scope of this work. Instead we look for evidence of the underlying relativistic physics in the density profile of a trapped gas, since this observable is universally available in experiments.   

\begin{figure}[htbp]
\begin{center}
   \includegraphics[width=\columnwidth]{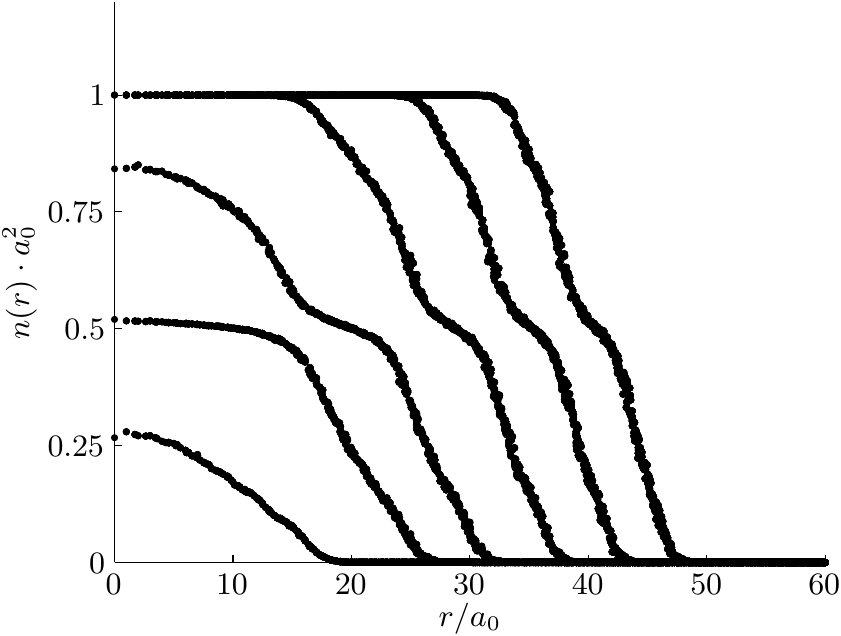}
\caption{Density profiles of noninteracting fermions in
  the combined lattice and trapping potential at zero temperature. From the left
  to the right $N=100, 500, 1000, 2000, 3000$, and $4000$. The trap strength is $\kappa a_0^2=0.01J$.}
\label{Fig:Densities}
\end{center}
\end{figure}

For simplicity we concentrate on the density $n({\bf r})$ of a zero temperature, noninteracting Fermi gas with $N$ atoms, since this only entails summing over the probability distribution of the $N$ lowest eigenstates
\begin{equation}
n({\bf r}) = \sum_{n=1}^{N} |\psi_n({\bf r})|^2.
\end{equation}
An ideal Fermi gas is realized with a degenerate single-component (fully polarized) gas of ultracold fermionic atoms due to the suppression of $p$-wave collisions and the symmetry requirements imposed on the wavefunction of identical fermions by the Pauli principle. In Fig.~\ref{Fig:Densities} we plot the density at each lattice point as a function of the distance from the center of the trap. The density at distance $r$ from the trap center can also be written as 
\begin{equation}
n(r)=\int_{-\infty}^{E_{\rm{F}}}\rho(E,r)dE, 
\label{eq:dens_LDOS}
\end{equation}
where the Fermi energy $E_{\rm{F}}$ is fixed by the constraint $N=\int n(r)d^2r$. In the center of the trap a band insulator with unit-filling is formed at sufficiently high particle number. By comparing with Fig.~\ref{fig:LDOS} one sees that unit-filling at site $i$ requires $E_{\rm{F}}>3J+\frac{1}{2}\kappa r_i^2$ such that the integral in (\ref{eq:dens_LDOS}) is over the full DOS of the homogeneous lattice (displaced by the local oscillator energy). 

Based on a local density approximation for the fermionic density profile it has previously been suggested that the Dirac points emerge as a shoulder in the density at a radius corresponding to half-filling~\cite{Zhu2007}. This can be understood as the position in the trap, where the local Fermi energy $E_{\rm{F}}(r)=E_{\rm{F}}-\frac{1}{2}\kappa r^2$ crosses the Dirac point located at zero energy in the homogeneous spectrum, such that the integral in (\ref{eq:dens_LDOS}) covers exactly {\textit{half}} of the displaced homogeneous lattice DOS. 
This prediction is confirmed by our calculation using the single-particle eigenstates of the tight binding Hamiltonian. The density profiles plotted in Fig.~\ref{Fig:Densities} show the anticipated shoulder at half-filling.

\section{Conclusion}
\label{sec:Conclusion}

We have shown how a confining potential alters the spectrum of a single atom in a honeycomb lattice. Even though the eigenvalues of the tight binding Hamiltonian are significantly modified by increasing the strength of the trapping potential, the characteristic spectrum of the homogeneous honeycomb lattice survives locally in the trap, provided the confining potential varies over a length scale much larger than the extent of a unit cell. This means that it should be possible to observe graphene-like physics with cold atoms in a honeycomb optical lattice, and hence that this system can be used to implement a relativistic quantum simulator. 

We have studied the density profile of a single-component Fermi gas and shown that the Dirac points emerge as a shoulder at half-filling.  
In addition, the local density of states suggests that the massless Dirac quasiparticles can be directly manipulated by a local spectroscopic probe. However, additional calculations are needed to conclusively demonstrate that the local dynamics is governed by the Dirac equation.

The single-particle density of states was fully described by a combination of analytical and semi-classical arguments. Importantly, the numerically calculated spectrum was reproduced with striking accuracy by a local density approximation based on the density of states of the homogeneous honeycomb lattice. This implies that statistical mechanics calculations of many-body systems in the combined trap and lattice potential can be done without resorting to numerical diagonalization of the tight binding Hamiltonian, provided the trapping potential is slowly varying over the size of a unit cell. 

\begin{acknowledgments}
We are grateful to S{\o}ren Gammelmark for  preparing Fig.~\ref{Fig:Schematic}.
N. N. acknowledges financial support by the Danish Natural Science Research Council.
\end{acknowledgments}

\appendix

\end{document}